# Line tension and structure of smectic liquid crystal multilayers at the air-water interface

Lu Zou, Ji Wang, Prem Basnet, and Elizabeth K Mann

Department of Physics, Kent State University, Kent, OH 44242, USA

**Abstract**

At the air/water interface, 4′-8-alkyl[1,1′-biphenyl]-4-carbonitrile (8CB) domains with different thicknesses coexist in the same Langmuir film, as multiple bilayers on a monolayer. The edge dislocation at the domain boundary leads to line tension, which determines the domain shape and dynamics. By observing the domain relaxation process starting from small distortions, we find that the line tension λ is linearly dependent on the thickness difference ΔL between the coexisting phases in the film, $\lambda = (3.3 \pm 0.2)\, mN/m \bullet \Delta L$. Comparisons with theoretical treatments in the literature suggest that the edge dislocation at the boundary locates near the center of the film, which means that the 8CB multilayers are almost symmetric with respect to the air/water interface.

**Introduction**

Line tension plays an important role in two-dimensional phase-coexistence systems. The coexistence of two phases implies energy associated with the boundary. This energy per unit length, defined as line tension, is an analog of surface tension in three dimensional systems[1]. However, this line tension has proved difficult to model theoretically, since it originates in the net attractive forces between surface molecules in the complex medium of the interface. To our knowledge, only two such theories exist, both based on treating the film as a continuous elastic medium with a thickness change across the film boundary.   Both the boundary profile and energy are determined by minimizing the total energy due to compression, bending, and surface tension across the profile. One such theory considers line tension in bilayer lipid membranes[2], assuming that the film is symmetric (center line of the hydrocarbon tails flat).   The other considers edge dislocations in smectic-A liquid crystalline films[3,] including the case where the films lie between two different fluids, with two different interfacial tensions. This theory has been compared to experimental results in free-standing films of 8CB[4] [5.]

Cell membranes and vesicles [6] with multiple lipid components form two-dimensional objects embedded in three-dimensional systems in which coexisting phases with an associated line tension are observed. Similar domains are observed in quasi-two dimensional systems confined to a flat surface, such as Langmuir films (molecularly thin layers confined by molecular interactions to the air/water interface[7].)   In all such cases, line tension, along with such factors as the viscosity of the film and substrate, also controls the shape, size and especially dynamics of the domain. The fine control of composition, surface pressure, temperature, and substrate possible in a Langmuir film makes these films valuable model systems for line tension in any quasi-two dimension systems, including biomembranes.

In order to minimize the line energy, a stretched domain in a Langmuir film relaxes back to a circular shape, shown in Fig. 1. This process can thus be used to estimate the line tension experimentally [8,9,10,11]. However, we know of only one systematic study [8] of line tension as a function of any control parameter. Here we use similar methods of deducing line tension between 8CB Langmuir multilayers of different thickness, as a function of the jump in that film thickness.

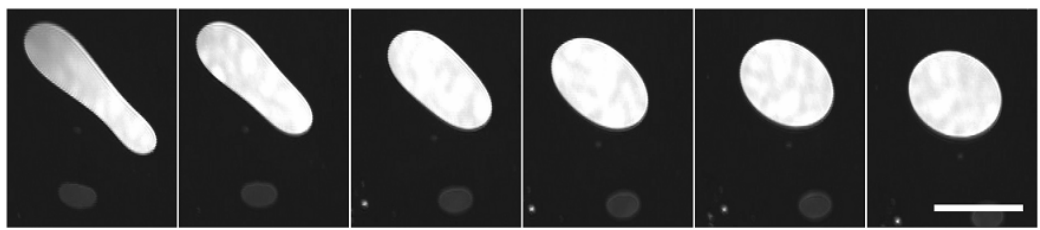

**Fig. 1. Brewster Angle Microscope images of a relaxation process. The bright domains are 8CB multilayers. The dark background is 8CB trilayer. The time interval between images is 1.0 sec. The white bar corresponds to 1.0 mm**

In bulk, 8CB forms a smectic-A liquid crystal, where ordered layers have their optical axis along the layer normal, in the temperature range of 21.5°C-33.5°C. It is thus unsurprising that at the air/water interface, 8CB forms a smectic-A film consisting of a multilayer structure in which all the layers are parallel to the interface. These multilayers are thought to consist of a monolayer at the water surface, with an integer number of interdigitated bilayers on top of it [12,13,14,15]. At room temperature, the trilayer begins forming at surface pressures of ~6.5 $mN/m$ [15]. As the surface layer is further compressed, the trilayer fills the surface. At this point, instead of an orderly first-order transition to an equilibrium five-layer state, domains with many different thicknesses form on the surface (Fig. 2.). The coexistence of domains with different thicknesses implies that there is an edge dislocation at the domain boundary with the corresponding line energy per unit length, or line tension. The fact that 8CB multilayers with different thicknesses coexist with the trilayer under the same external physical conditions makes it possible for us to measure this line tension as a function of the jump in thickness at room temperature. We will consider both the origin of this line tension and its implications on the structure of the film at the boundary.

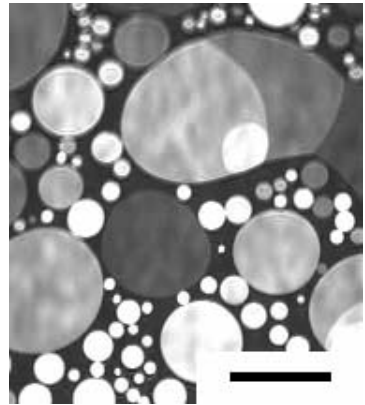

**Fig. 2. Brewster Angle Microscope image of the coexistence of 8CB multilayers. The background is 8CB trilayer. The layer reflectivity increases with thickness, so that different grey levels correspond to different thicknesses. The rippled variation in color within one domain is due to variations in illumination. The black scale bar is 1.0 mm.**

In order to measure the line tensions from hydrodynamic relaxation, we needed to find a way to produce isolated domains to avoid hydrodynamic forces between the domains.  To estimate the line tension, we determined the characteristic relaxation time of deformed domains towards a circle, and analyzed the result with a standard expression for the relaxation of small deformations in two-dimensional fluid domains.

The reflectivity allowed us to estimate the domain thickness, and in particular, the thickness jump (or Burgers vector $b$[16]). We finally consider the implications of the line tension vs. Burgers vector expression in terms of the structure at the boundary line.

## Materials and Experimental Methods

### Materials and experimental setup

Commercial 8CB (99%, Fisher), further purified by chromatography technique, was dissolved in hexane (OPTIMA, Fisher) to obtain an 8CB solution at ~0.3 mg/ml. A PurelabPlus/UV system produced the pure water for the substrate (resistivity >18.2MΩ/cm; passes shake test: i.e., small bubbles break as they reach the surface.). Solution was deposited on the pure water surface in a well-cleaned Langmuir trough (mini-trough, KSV). The hexane evaporated, leaving an 8CB monolayer or multilayer behind. A pair of symmetric movable barriers controlled the water surface area, and thus the surface concentration. The layers were imaged with a standard Brewster Angle Microscope [17]. A polarized red laser beam (668 *nm*, SDL 7470-P6) was incident to the surface at Brewster angle. The reflected beam was captured by a CCD camera (GPMF602, Panasonic 768x494 pixels) and finally recorded by a computer with image grabber card at the rate of 30 frames/second. The temperature was controlled to be 18°C with 70*%* humidity. A Wilhelmy plate was used to monitor the surface pressure. The surface potential was measured with a KSV SPOT1 surface potential meter. A very thin platinum wire was used to stretch 8CB domains by inducing shear in the underlying fluid. Both the Wilhelmy plate and platinum wire were soaked in an ethanol/KOH solution and rinsed with pure water as described above.

**Forming isolated domains**

Bulk 8CB transforms from crystal to smectic-A at 21.5°C. However, at the air-water interface, 8CB molecules form a stable smectic-A structure when the temperature is even lower than this. Previous works [12,15] show that, when the surface is compressed, 8CB monolayer collapses to trilayer domains quickly, with the new-forming trilayer distributed everywhere on the surface at the same time. The monotonic surface pressure (Fig. 3) also demonstrates the absence of a significant nucleation barrier. The surface potential in the monolayer region shows the random values between those of monolayer and gas phases typical of large domains, but it plateaus once the trilayer begins forming. This implies that there is no significant difference in the dipole density between a monolayer and a trilayer, and by extension any number of additional bilayers, as expected given the bilayer symmetry. The optical isotropy of 8CB monolayer and multilayers was tested by changing the orientation of an analyzer through which the reflected beam arrives on the CCD camera; no sign of anisotropy was observed. This is as expected both from previous measurements [15] and from the nearby bulk smectic-A phase.

During compression, the whole surface becomes covered with trilayer, and then the trilayer collapses to thicker multilayer domains which include different integer numbers of bilayers (Fig. 2.). The thicker domains appear randomly everywhere on the top of trilayer, as the trilayer domains do on the monolayer. This coexistence state of trilayer and multilayers is stable over several hours. This property of 8CB makes it possible for us to study the coexistence of the trilayer and other different multilayers, although such multiple coexisting layers should represent a metastable rather than a true equilibrium state. The collapse processes happen randomly and simultaneously everywhere on the trilayer (such as Fig. 2). Because of the viscosity of the substrate and the incompressibility of the film, the motion of the surrounding domains affects the relaxation process significantly. In order to reduce this hydrodynamic effect, only the relaxation of those domains far apart from each others is analyzed. Therefore, a special compression/decompression routine, discovered by trial and error, was performed in order to obtain isolated domains with different thicknesses, such as the one shown in Fig. 1.

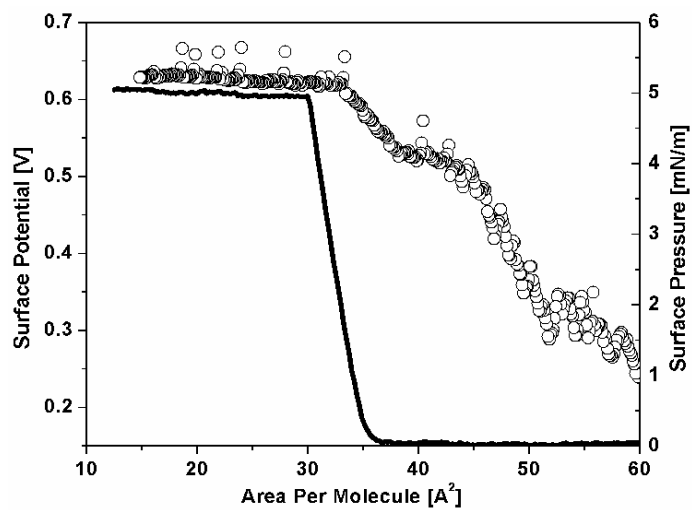

**Fig. 3. Surface pressure and surface potential of 8CB versus the area per molecule at air/water interface. Solid line: surface pressure; open circles: surface potential.**

The routine follows: (1) Deposit the appropriate amount of 8CB/hexane solution on the water surface with the barriers far from each other so that the surface is $\sim 90\%$ covered by 8CB monolayer after the hexane evaporates. (2) Move the barriers at the rate of $\sim 100 mm/\min$ to decrease the surface area and compress the 8CB layers until the whole surface is covered by a trilayer and some brighter (thicker) domains appear. Generally, at this step, the thicker domains are relatively small and close to each other, and thus unsuitable for relaxation measurements. (3) Stop the barriers for $\sim 5\min$ and then move the barriers back at $\sim 10 mm/\min$ to decompress the 8CB layer. While the barriers are moving back, the brighter domains disappear first and then holes, which are monolayer, open in the trilayer film. (4) Stop the barriers when the trilayer is net-like on the top of monolayer at the surface. Leave the barriers for about half an hour to let the trilayer net relax into isolated domains. (5) Then, compress the 8CB film at $\sim 10 mm/\min$. The trilayer covers the whole surface again and finally some thicker and isolated domains appear, including some that are big and isolated enough for the relaxation measurements. Sometimes, the thicker domains are too close to each other or the sizes are too small; one can adjust this by moving the barriers back and forth at $\sim 10 mm/\min$. One can also use a controlled air stream to move the domains in or out of the field of view.

    Note that in order to form large isolated domains, it was necessary to allow large isolated trilayer domains to form on the monolayer. It appears that the formation of multilayers has some memory of the proceeding trilayer. However, that trilayer remains fluid, and we observed no defects, so that we do not understand this observation.

**Determining the line tension from the relaxation process**

Several different methods exist to stretch the domain [9][10]. Here, we apply a very direct one, by inducing shear in the substrate, which thus shears the film. Once a desired domain is ready on the surface, one can use the tip of a very thin, carefully cleaned as discussed above, platinum wire (diameter = 1.3 mm) to stir the film near the domain in order to stretch it. While this method is rather rough and ready, we have found it more successful than more controlled methods, such as the four-roll mill [10], in keeping isolated domains of many different thicknesses for the 8CB multilayer system. To obtain a good relaxation process, one has to control the stirring speed very well. Too high a stirring speed may cause excessive subfluid flow and the whole domain may move out of the field of view while relaxing, while too low a speed may not stretch the domain far enough for us to collect adequate data for analysis. Furthermore, after the wire leaves the surface, it takes a couple of seconds for subfluid flow to calm down, as evidenced by the general surface flow. The relaxation process during this time is not counted in the analysis. In addition, domains with different thicknesses and different sizes have different physical properties, and thus require different stresses to stretch them appropriately. Here, the speed of the manual stirring cannot be readily quantified. The four-roll mill provides better control on the domain stretching process, and is thus often preferable [18].

The whole relaxation process is recorded by a computer in a video file. We use "*ULead VideoStudio 7.0*" to capture a series of individual images from the video with a real time scale. To analyze the relaxation process, one defines a distortion $\Theta$ by $\Theta = L/W - 1$, where $L$ and $W$ are the length and the width of an elliptically deformed domain [9]. Then, by measuring $L$ and $W$, the distortion $\Theta$ is calculated for each image. The characteristic time $T_c$ is determined from the first order exponential decay fitting result of the plot $\Theta$ vs. time. Finally, the line tension $\lambda$ is determined from $T_c$ and the corresponding film thickness of the domain is determined quantitatively from its brightness, as discussed below.

**Estimating the film thickness**

The reflectivity $R = I_r / I_i$, with $I_r$ the reflected and $I_i$ the incident intensity, of the thin film depends on the thickness. To first order in the ratio of the film thickness $D$ to the wavelength of the light [19], $R \propto D^2$. This is valid whether the optic axis of the film is perpendicular to the surface or not; here the optical axis is perpendicular to the surface, so that all domains of the same thickness have the same reflectivity. In our experiment, the gray level of CCD images is given by a number in the range of 0 to 255, corresponding to optically saturated and totally dark states of the camera. Due to the limitation of our experimental environment, a non-zero gray level value, $G_o$, always exists as a dim background. In all the experimental results used in this work, $G_o$ is very stable with a relative fluctuation less than 1%. The grey level was found to be linear in the intensity within 2 percent, as determined by verifying Malus' law as a function of polarizer orientation [20]. Thus, if $G_m$ is the gray level value for the domain with m layers, $G_o - G_m$ is proportional to the intensity of the reflected beam, $I_m$. Then, the relative reflectivity $R_m$ can be expressed as

$$R_m = \frac{I_m}{I_1} = \frac{G_o - G_m}{G_o - G_1} \qquad (1)$$

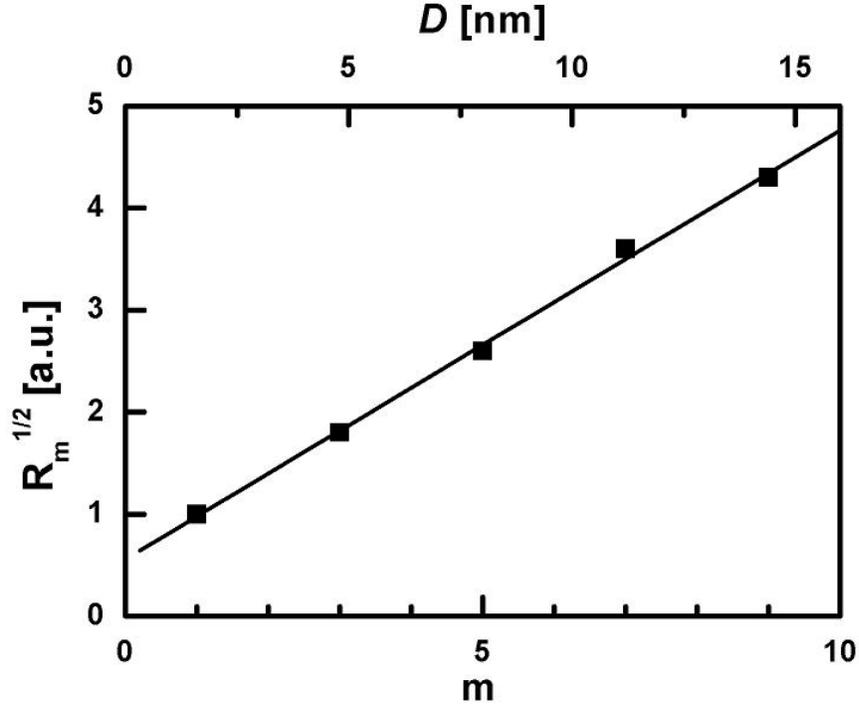

**Fig. 4. The dependence of the relative reflectivity ($R_m$) on the number of 8CB layers ($m$) and the film thickness ($D$). The solid squares correspond to the data from the measurement in reference [20]. The film thickness $D$ in nm is estimated based on X-ray measurements of films on solids [21] (see text).**

Here, the relative reflectivity $R_1$ for monolayer is set as 1. The relation between $R_m$ and the number of 8CB layers $m$ was determined from experimental results in reference [20]:

$$R_m^{1/2} = a + b \bullet m \qquad (2)$$

where $a = 0.56 \pm 0.07$ and $b = 0.42 \pm 0.01$. By determining the relative reflectivity of two different layers, we thus can convert directly from the relative reflectivity to the number of bilayers $(m-1)/2$ with Equation (2). Using the monolayer and bilayer thicknesses as determined by X-ray reflectivity of 8CB on silicon wafers [21], $d_{mono} = (1.2 \pm 0.1)nm$ and $d_{bi} = (3.3 \pm 0.1)nm$, the number of layers can be converted into an estimated film thickness *D*. Therefore, the real thicknesses for different 8CB multilayers can be estimated, as shown in Fig. 4.

**Result**

Several groups have used the relaxation of domains towards a circle to determine line tensions [8,9,10]. Here, we use the relaxation from relatively small distortions, namely $\Theta \leqslant 1$, for which distortions relax exponentially in time *t* as [9,22]

$$\Theta = \left(\frac{L}{W} - 1\right) \propto \exp(-\frac{t}{T_c}) \tag{3}$$

where $T_c$ is a characteristic relaxation time that depends on the line tension and the viscosity of both substrate and film. By assuming incompressibility and constant viscosity in the surface layer, Stone and McConnell [22] gave a general result for $T_c$. The electrostatic repulsion due to the alignment of molecular dipoles at the surface can play an important role in some systems [8], but not in this one, since within each bilayer the molecular dipoles in opposite direction cancel each other. While both surface and bulk viscosity can exist in principle, dissipation in the surface can be neglected if $\eta_s \ll \eta_b R$, where $\eta_b$ is the viscosity of the substrate, $\eta_s$ is the surface viscosity of the film and *R* is the characteristic size of the domain. It is difficult to estimate surface viscosities except from difficult, direct measurements, since the ordering and compression in a surface layer can lead to surface viscosities orders of magnitude larger than the naïve estimate of the measured bulk viscosity multiplied by the thickness of the layer, here $\eta_{8CB} D$. In the 8CB case, the multilayer is directly analogous to a smectic-A phase, so that viscosity measurements on this phase should also apply to the monolayer. The ratio of bulk viscosities of 8CB and pure water, i.e. $\eta_{8CB}/\eta_{water}$, is $<10^2$ [23] for smectic-A phase at room temperature, depending on the time scale and the layer orientation. The thickness $D$ of the 8CB layer is $<3\times10^{-8}m$. Thus we estimate $\eta_s/\eta_b \sim \eta_{8CB}D/\eta_{water} <10^{-6}m$. Since the domain size $R \sim 10^{-4}m$, we indeed see that the surface viscosity can be neglected.

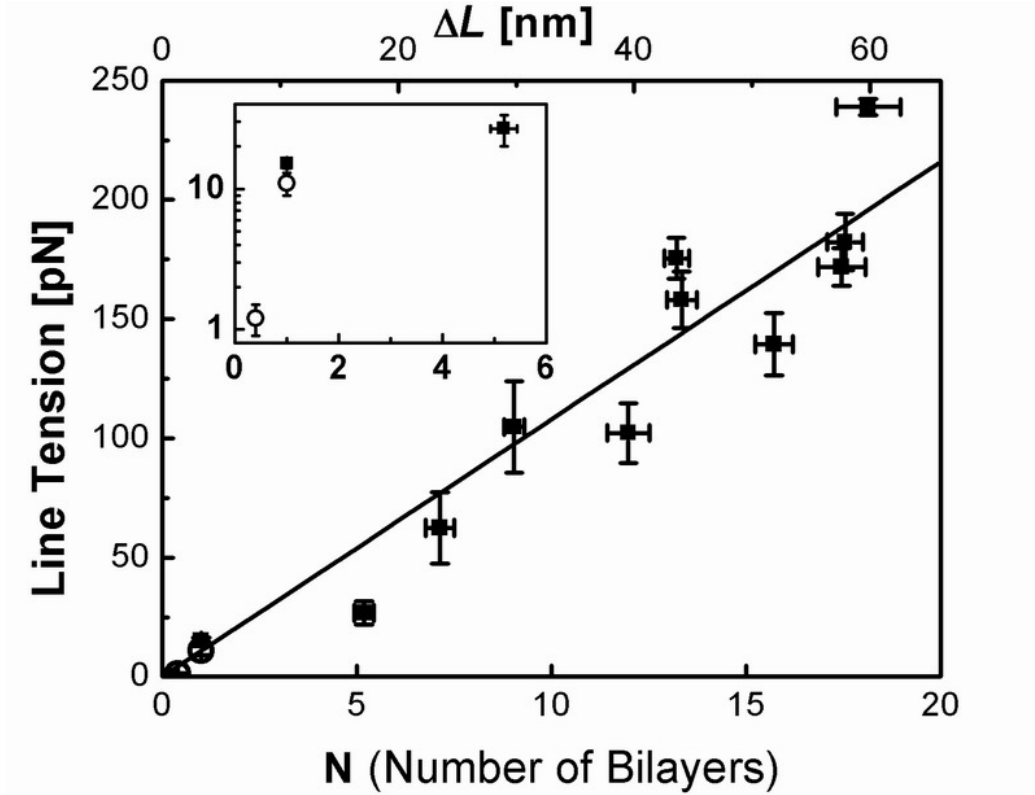

**Fig. 5. Plot of line tension vs. thickness jump in number of bilayers and nanometers. Empty circles are from Lauger et al. [10].**

Under the limitation of $\eta_s << \eta_b R$ and eliminating the competing effect of electrostatic repulsion, the relaxation rate found in reference [22] can be written as [9]

$$T_c = \frac{\eta_b R^2}{\lambda}\frac{(2n+1)(2n-1)\pi}{4n^2(n^2-1)} \qquad (4)$$

where $\lambda$ is the line tension of the domain, $n$ is the mode number of a very small distortion, and $R$ is the radius of the round domain after relaxation. For an ellipsoidal distortion, $n = 2$,

$$T_c = \frac{5\pi}{16} \frac{\eta_b R^2}{\lambda} \qquad (5)$$

By following the compression/decompression routine described in the previous section, we can obtain isolated domains appropriate for analysis. For the case of small distortions, Equation (3) and (5) are consistent with our experiment data: the distortion $\Theta$ exponentially decayed with time for $\Theta < 1$, and for domains of different sizes but with the same thickness, the characteristic time $T_c$, which would have been linear in $R$ if surface viscosity played an important role, depended linearly on the domain size $R^2$ within experimental error.

Thus, the relaxation time $T_c$ allows us to deduce the line tension from Equation (5). This line tension is shown for domains of different thickness in Fig. 5. Lauger and coworkers [10] applied Equation (5) to find the line tensions for the 8CB monolayer and the 8CB bilayer on the top of the monolayer as $\lambda = (1.2 \pm 0.3) \times 10^{-12} N$ and $\lambda = (1.1 \pm 0.2) \times 10^{-11} N$, respectively. We repeated the measurement of the bilayer on the monolayer to find $\lambda = (1.5 \pm 0.1) \times 10^{-11} N$, which is within experimental uncertainty of Lauger's result. The line tension for a monolayer in a gas background is clearly anomalous, but for any number of bilayers, the line tension is linear in the number of bilayers. If one defines $\Delta L$ as the thickness difference between the thicker 8CB domain and the substrate 8CB layer (the substrate layer is trilayer in all experiments here), we find $\lambda / \Delta L = (3.3 \pm 0.2) mN/m$ (Fig. 5).

**Discussion and Conclusion**

A first estimate of the expected line tension would be to model the line as an abrupt transition from one layer thickness to the other, and consider the excess surface energy at that jump, $\gamma \Delta L$. Estimating the surface tension at ~30mN/m, this would lead to $\lambda = 30 mN/m \bullet \Delta L$, about an order of magnitude bigger than what we measure. Such an abrupt boundary is thus unlikely, and one should consider a gentler curve for the boundary profile, which would minimize surface energy at the expense of the elastic energy needed to compress and bend the layers.

Both the behavior of the line tension as a $\Delta L$. and the value of $\lambda/\Delta L$. are very comparable to that seen in thick 8CB free-standing films, where $\lambda/\Delta L = 2.5 mN/m$ at 22°C, which extrapolates, with the methods of that article [5], to $2.9 mN/m$ at 18°C.

In 1991, Lejcek and Oswald [3] developed a general expression for the energy of edge dislocations in smectic-A liquid crystal films including the elastic interaction energy of edge dislocations with surfaces, by finding the boundary profile necessary to minimize the combination of surface and elastic energies. For a symmetric free-standing smectic-A film in air [4], the line tension of the edge dislocation is reported to be proportional to the Burgers vector $b$ [16], , as discussed above. To our knowledge, there have been no direct measurements about how the Burgers vector affects the elastic energy of the edge dislocation in a Langmuir film. In the case of the coexistence of 8CB multilayers and trilayer, the Burgers vector $b$ is equal to the thickness difference $\Delta L$, assuming a single dislocation line. Our experimental result shows a line tension, reflecting the elastic energy of the dislocation, proportional to the Burgers vector. In principle, the boundary might be split into two dislocation lines for very large $\Delta L$ [4], which would change the boundary energy.

Previous theory [3] then gives an expression for the equilibrium position of the edge dislocation inside a general film with two different surfaces.

$$t = z_0/D = \frac{1}{1+Q} \quad \text{with} \quad Q = \left(\frac{A_1}{A_2}\right)^{2/3} \tag{6}$$

Here, $D$ is the thickness of the film and $z_0$ is the position of the dislocation as $z_0 = 0$ is at the film/water interface. The parameter $A = \left(\frac{\gamma - \sqrt{KB}}{\gamma + \sqrt{KB}}\right)$; here, $\gamma$ is the surface tension, $K$ is the curvature constant and $B$ is the elastic modulus of the layers. The subscripts 1 and 2 correspond to the air/film and the film/water interfaces respectively.

For 8CB layers, $K = (5.2 \pm 0.3) \times 10^{-12} N$, $B = 1.63 \times 10^7 N/m^2$ at 18°C [5], and $\gamma_1 \cong 28.46\, mN/m$ [24]. We estimate the surface energy of the 8CB/water interface by considering that the surface energy per unit area is given by $\gamma = \gamma_1 + \gamma_2 + U$, where $U$ is the interaction energy between the two interfaces. The Hamaker constant quantifying van der Waals interactions across this surface is unknown, but about 10 times less than it would be for a free-standing film [25], so that ignoring U should give a good order of magnitude estimate for the relatively large surface tensions observed. The surface tension is $\gamma = \gamma_{water} - \pi = 72.8\,mN/- 6.5\,mN/m = 66.3\,mN/m$ so that, ignoring U, $\gamma_2 \cong 37.8\,mN/m$. With this estimate, $A_1 = 0.51$ and $A_2 = 0.61$. Thus, $t \cong 0.53$. This implies that the profile of the 8CB layer at the air/water interface is approximately symmetric with respect to upper and lower interfaces, as shown in Fig.

6. The edge dislocation lies near the midplane of the 8CB layer. Usual schematics of the Langmuir film profile shows the water surface as a horizontal plane, with the Langmuir layer above it. The present picture is, however, consistent with the picture of a drop of one fluid on another [7], except that the angles are determined partially by the elasticity in the film.

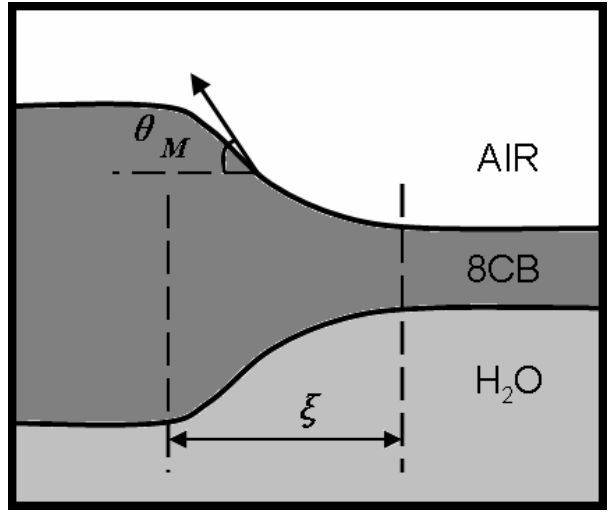

**Fig. 6. Schematics of the profile of 8CB layer at the air/water interface. $\xi$ is the dislocation width and $\theta_M$ is the maximum tilt angle on the boundary.**

Within this model, the theoretical prediction [26] for the associated line tension is

$$\lambda_\infty = 2\sqrt{B\Lambda\gamma_c b} \qquad (7)$$

Here, $\lambda_\infty$ denotes the line tension of an edge dislocation in an infinite medium. Our experimental result, $\lambda_\infty/b = (3.3 \pm 0.2) mN/m$, suggests a cut-off energy $\gamma_c = (0.27 \pm 0.03) mN/m$. A similar but slightly lower value was found for free-standing smectic-A films [4], of the same material at a higher temperature of $21°C$, $\gamma_c = 0.14 mN/m$.

The profile of the film is, within this theory, determined by a balance of surface tension and elasticity. For a perfectly symmetric film, and this one should be nearly so, the profile of the 8CB film at air/water interface can be characterized by the dislocation width $\xi$ and maximum tilt angle $\theta_M$ as shown in Fig. 6:

$$\xi \approx 2\sqrt{\Lambda D} \quad and \quad \theta_M \approx \frac{b(1-A)}{\xi\sqrt{2\pi}} \tag{8}$$

For the case of one bilayer on top of a trilayer (i.e., $N=1$), we can calculate that $\xi = 7.9nm$ and $\theta_M = 5^o$. For $N=18$, the thickest film observed here, $\xi = 22.6nm$ and $\theta_M = 30^o$. From the isotherm areas per molecule for monolayer and trilayer, the area per molecule pair in the bilayer is $0.23nm^2$, which implies that the 8CB molecular dimension in the bilayer is $\sim 0.48nm$. Thus, the dislocation width is between 15 and ~50 8CB molecular widths for the films studied here. Both the dislocation width and maximum tilt angle increase with film thickness. Note that using a single parameter to characterize the line tension assumes that the profile can continuously adjust to its equilibrium value within the time scale of the relaxation; this assumption is consistent with our experimental results.

In conclusion, we studied the coexistence state of 8CB films with different thicknesses at the air/water interface at room temperature in this work. By analyzing the relaxation processes for different 8CB domains, we found that the line tension depends linearly on the thickness jump with a slope of $(3.3 \pm 0.2)mN/m$. This is, for the first time for a Langmuir monolayer, in reasonable agreement with an existing theoretical treatment of line tension. This treatment, balancing the extra surface energy and elastic energy due to the line defect, was developed for free standing smectic-A films but applicable to Langmuir films as well. This treatment also suggests that the profile of the 8CB layer at the air/water interface is approximately symmetric with respect to upper and lower interfaces and that the edge dislocation lies almost in the midplane of the 8CB layer. The estimated width of the line ranges from about 15 to 50 molecular dimensions, increasing with the jump in layer thickness.

Note that the line tension of a monolayer in a gaseous background is about 3 times smaller than would be predicted by an extrapolation from the multilayer results. Previous authors, who looked at the monolayer and trilayer cases, suggested that the origin of this anomaly is much large dipole moment density contrast between the monolayer and gas phases, leading to much larger long-range electrostatic repulsion renormalizing the line tension. Another possibility is that interactions with the water substrate significantly decrease the attractive interactions in the first monolayer, which would affect subsequent bilayers significantly less. Different molecular orientations in monolayer and thicker layers could lead to this result. Sufficiently accurate [18] measurements of the line tension as a function of size in the monolayer regime should be able to distinguish the two possibilities.

## Acknowledgment

This material is partially based upon work supported by the National Science Foundation under Grant No.9984304. We also acknowledge Dr. J.V. Selinger for a very helpful discussion.